\documentclass[12pt]{article} 

\usepackage{epsfig}
\usepackage{graphicx}
\usepackage{amsmath}
\usepackage{amssymb}
\usepackage{url}

\begin{document}

\title{Kinematics effects of atmospheric friction in spacecraft flybys}

\author{L. Acedo\thanks{E-mail: luiacrod@imm.upv.es}\\
Instituto Universitario de Matem\'atica Multidisciplinar,\\
Building 8G, $2^{\mathrm{o}}$ Floor, Camino de Vera,\\
Universitat Polit$\grave{\mbox{e}}$cnica de Val$\grave{\mbox{e}}$ncia,\\
Valencia, Spain\\
}

\maketitle

\begin{abstract}
Gravity assist manoeuvres are one of the most succesful techniques in astrodynamics. In these trajectories the spacecraft
comes very close to the surface of the Earth, or other Solar system planets or moons, and, as a consequence, it experiences
the effect of atmospheric friction by the outer layers of the Earth's atmosphere or ionosphere. 
In this paper we analyze a standard atmospheric model to estimate the density profile during the two Galileo flybys, the NEAR and the Juno flyby. We show that, even allowing for a margin of uncertainty in the spacecraft cross-section and the drag coefficient, the observed $-8$ mm/sec anomalous velocity decrease during the second Galileo flyby of December, 8th, 1992 cannot be attributed
only to atmospheric friction. On the other hand, for perigees on the border between the termosphere and the exosphere the friction only accounts for a fraction of a millimeter per second in the final asymptotic velocity.
\end{abstract}

{\bf Keywords:} Spacecraft flyby, Atmospheric friction, Perturbation theory, Flyby anomaly

\section{Introduction}
\label{intro}

In the sixties of the past century \cite{Flandro,Transfer} the aerospace engineer G. A. Flandro devised an ingenious proposal for a spacecraft to extract energy from the gravitational field of a planet using a gravity assist or flyby manoeuver. In his idea the planet is 
seen as a field of force moving relative to the inertial heliocentric or barycentric coordinate system so it can transfer
a certain amount of kinetic energy to the passing spacecraft with respect to that inertial system. The transfer can be positive
or negative depending upon the particular geometry of the flyby but its main objective was to reduce the required launch energy and the time for arriving at a given destination.

Flandro also foresaw the fact that all major planets from Jupiter to Neptune would be almost aligned on the same side of the Sun during the decades of the seventies and the eighties allowing for the design of a ``grand tour'' project in which a spacecraft should successively explore Jupiter, Saturn, Uranus and Neptune in a period of twelve years \cite{Flandro}. This project was effectively carried
out by the Voyager 1 and Voyager 2 missions launched on September, 5th and August, 20th, 1977, respectively \cite{Butrica}. The flybys of Jupiter gave all the necessary energy boost to travel to Saturn in less than four years from the launching time. Moreover, these manoeuvers also provided an excellent opportunity to take close images of the giant planets as they have never been seen before.

Since then, flybys have become an integral part of space exploration and, in particular, the Juno spacecraft, now in orbit around Jupiter, is programmed to perform a total of 37 close flybys of the planet to gather information about its atmosphere and magnetic field \cite{JunoMissionI,JunoMissionII,JunoMissionIII}. Moreover, many missions have included flybys of Venus and the Earth in his way to bodies in the outer Solar System. It was in the analysis of the trajectories of these flybys around the Earth that a team lead by Anderson discovered
an unexpected velocity change in the Galileo flyby of Earth that took place on December, 8th, 1990 \cite{Anderson2008}. Fitting the post-encounter
residuals for Doppler data they found that a small value, to be interpreted as a velocity increase of $3.92$ mm$/$sec, cannot be
attributed to any known perturbing effects considered in the orbit determination program. The anomaly also appeared in the ranging data, so it cannot be dismissed as a systematic error in the Doppler tracking method. 

Two years later the same spacecraft performed a flyby with a perigee at an altitude of $303$ km over the Earth surface so, it 
crossed the middle layers of the termosphere were a detectable perturbation by atmospheric friction can be measured. In fact,
a decrease of $8$ mm$/$sec was found after taken into account all other sources of perturbations. However, Anderson et al. \cite{Anderson2008} claimed that only a $3.4$ mm$/$sec velocity decrease could be explained as a consequence of friction. Similar flyby anomalies were found in subsequent flybys by the NEAR, Cassini, Rosetta and Messenger spacecraft but, apparently, they have not arisen in the most recent Juno flyby \cite{Thompson} and in the 2007 and 2009 flybys by Rosetta \cite{Jouannic}.

The problem has attracted an interdisciplinary interest from aerospace engineers to physicists. From the point of view
of classical physics, there have been many attempts for an  explanation: L\"ammerzahl et al. \cite{LPDSolarSystem} classified and analyzed
several sources of classical perturbations according to their order of magnitude including atmospheric drag, ocean and solid
Earth tides, spacecraft's electric charge, magnetic moment, Earth albedo, Solar wind and spin-rotation coupling. Their preliminary assessment showed that all these effects were very small to account for the anomalies. Other studies have analyzed in 
detail the General Relativistic corrections on hyperbolic orbits \cite{IorioSRE2009}, anisotropic thermal emission off the spacecraft and thermal radiation pressure \cite{Rievers2011}, Lorentz
acceleration of a charged spacecraft \cite{Atchison}, the Lense-Thirring effect for hyperbolic orbits \cite{Hackmann}, and the effect of ocean tides and tesseral harmonics \cite{AcedoMNRAS}. All these studies have shown that no definitive explanation can be obtained from these classical approaches.

As the possible explanations within standard physics are being analyzed in detail and dismissed we have also an increasing 
number of proposals, more or less well-motivated, invoking non-standard physics such as the existence of a dark matter halo
surrounding the Earth and whose interactions with the spacecraft could account for the anomalous accelerations as proposed
by Adler \cite{Adler2010,Adler2011}. Other models include a modification of Newtonian potential \cite{Nyambuya2008,Lewis2009,
Wilhelm2015,Bertolami2016}, some extensions or modifications of General Relativity \cite{GerrardSumner,Varieschi2014,Acedo2014,Acedo2016,Hafele,Pinheiro2014,Pinheiro2016}, violation of
established principles such as the principle of equivalence \cite{McCulloch} or Lorenz invariance \cite{Cahill}, and phenomenological formulas \cite{Busack}. Nevertheless, these  models have not provided a satisfactory explanation of all the anomalies and, in particular, they are not able to explain
the null cases: the absence of extra corrections to the asymptotic velocity of the spacecraft in the case of the Juno flyby in 2013 \cite{Thompson} and the two Rosetta flybys of 2007 and 2009 \cite{Jouannic}. In this situation of impasse it is becoming increasingly important to
elucidate any classical contributions to the flyby anomalies to determine if they are really new phenomena deserving further study or they are just the result of a missing, but already known to physics, element in the orbit determination program.

On the other hand, flybys inside planetary atmospheres with the objective of scientific research are more likely to be 
programmed in future missions. The Juno spacecraft is currently performing successive flybys of Jupiter at a minimum altitude
around $4000$ km over the upper clouds of the giant planet \cite{JunoMissionI,JunoMissionII,JunoMissionIII}. This zone lies in the exosphere of Jupiter and some friction 
effect can be expected \cite{Miller2005}. In this paper we will study a simple atmospheric model to calibrate the density profile of the
Earth in terms of the solar radio flux at $10.7$ cm, the $F10.7$ index, and the geomagnetic index, $A_p$. The solar radio
flux at $10.7$ cm (or $2800$ MHz) is an excellent indicator of Solar activity and the geomagnetic index measures the general
level of geomagnetic activity over the globe for a given day and it correlates with Solar storms and the number of sunspots
\cite{UKSolar,KingHele}.
During solar storms the ionosphere expands its zone of influence and higher densities are achieved at the same altitude in
comparison with quieter days. By using this atmospheric model, and the dimensions of the spacecraft, we will estimate the friction deceleration and its kinematic effect on the orbit during the period in which the spacecraft lies at an altitude
lower than $1200$ km, where we can consider that the atmosphere merges with interplanetary space.

The results will show that only for the second Galileo flyby of Earth a measurable impact on the trajectory is obtained.
In this case the expected decrease of the asymptotic velocity would be $-3.68(99)$ mm$/$sec, which coincides with the estimation
of Anderson and his team. The error comes from the uncertainties in the effective area of the spacecraft during the flyby but
the result, certainly, cannot be larger in magnitude that the observed $-8$ mm$/$sec decrease and we must conclude that a residual anomaly of $-4.32(99)$ mm$/$sec is still unexplained. In the other flybys with perigees at $500$ km over the Earth surface and above we obtain only minor corrections of a fraction of mm$/$sec which do not seriously change the results for
the residual anomaly.

The paper is organized as follows: In section \ref{model} we discuss the ionosphere's model and the expression for the friction 
force and we provide the parameters for the Galileo, NEAR and Juno spacecraft. In section  \ref{pertur} we give the equations
of motion for the perturbations and the parameters corresponding to the osculating keplerian orbit at perigee. The integration
results are obtained in Section \ref{results}. Finally, in Section \ref{conclusions} we briefly discuss our results in connection 
to the flyby anomaly and its possible application to other planetary atmospheres.

\section{Atmospheric model for the thermosphere and spacecraft's friction}
\label{model}

The ionosphere is a part of the Earth's atmosphere that extends from altitudes ranging from $60$ km to $1000$ km. It includes the upper part of the mesosphere, the termosphere and the exosphere. Its name comes from the fact that it is composed mainly by
ionized atoms and molecules; the ionization takes place as a consequence of the ultraviolet radiation from the Sun and its 
structure is also greatly affected by Solar storms as particles emitted in Coronal mass ejections reach the Earth's magnetosphere
and precipitate from there into the termosphere, heating it, and changing its density \cite{KingHele}.

Lower termosphere is also the region in which aurora borealis are formed and it is a region of very low density but it still
causes important consequences on the orbits of satellites and spacecraft perfoming flyby manoeuvres. As a consequence of friction the satellites
suffer orbital decay into the mesosphere and they finally disintegrate as a consequence of the thermal effects of friction in
the higher density layers of the atmosphere \cite{KingHele}. The basic model for the thermosphere starts by defining a temperature as follows:
\begin{equation}
\label{temp}
T=900+2.5 (F10.7-70)+1.5 A_p\; \mbox{ Kelvin}\; ,
\end{equation}
where $F10.7$ is the solar radio flux at $10.7$ cm and $A_p$ is the geomagnetic index. The solar radio flux is measured
in solar flux units, or SFUs, with one SFU corresponding to $10^{-22} \mbox{ Watts/m$^2$ Hz}$ and it usually ranges between $65$ SFUs, when the Sun is quieter, to $300$ SFUs in cases of strong solar storms. The geomagnetic index ranges from $0$ to $400$. We also define a molecular mass as a function of altitude:
\begin{equation}
\label{mol}
m=27-0.012(h-200)\; ,
\end{equation}
where $h$ is the altitude from the Earth surface measured in km. The model is used for $180 \mbox{ km} < h < 500 \mbox{ km}$ but
we will extend it a bit further to $h \simeq 1000 \mbox{ km}$ to estimate the impact of friction on the NEAR and Juno trajectories.

From Eqs. (\ref{temp}) and (\ref{mol}) we obtain the following density profile:
\begin{equation}
\label{profile}
\rho=6\times 10^{-10} \, e^{-(h-175)m/T} \; \mbox{kg$/$m$^3$}\; ,
\end{equation}
Notice that the density at a given altitude, $h$, increases in periods of strong solar activity because both the solar radio
flux, $F10.7$, and the geomagnetic index, $A_p$, become larger and the thermosphere is heated, i.e., we obtain a larger value
of $T$ from Eq. (\ref{temp}). At table \ref{tab1} we list the altitude at perigee and the $F10.7$ and $A_p$ indexes. These were
obtained from the UK Solar System Data Centre \cite{UKSolar}.

\begin{table*}
\caption{Altitude at perigee, solar radio flux, geomagnetic index and spacecraft mass during the flybys. \label{tab1}}
\centering
\resizebox{\textwidth}{!}{
 \begin{tabular}{lccccc}
  Spacecraft & Date & Altitude (Perigee) in km & $F10.7$ (SFU) & $A_p$  & mass (kg) \\
  \noalign{\smallskip}
  Galileo I & 12/8/1990 & 960 & 230 & 8 & 2497 \\
  \noalign{\smallskip}
  Galileo II & 12/8/1992 & 303 & 129 & 26 & 2497 \\
  \noalign{\smallskip}
  NEAR & 1/23/1998 & 539 & 97 & 4 & 730 \\
  \noalign{\smallskip}
  Juno & 10/9/2013 & 559 & 113 & 29 & 3625
  \end{tabular}}
\end{table*}

\begin{figure}
\includegraphics[width=\columnwidth]{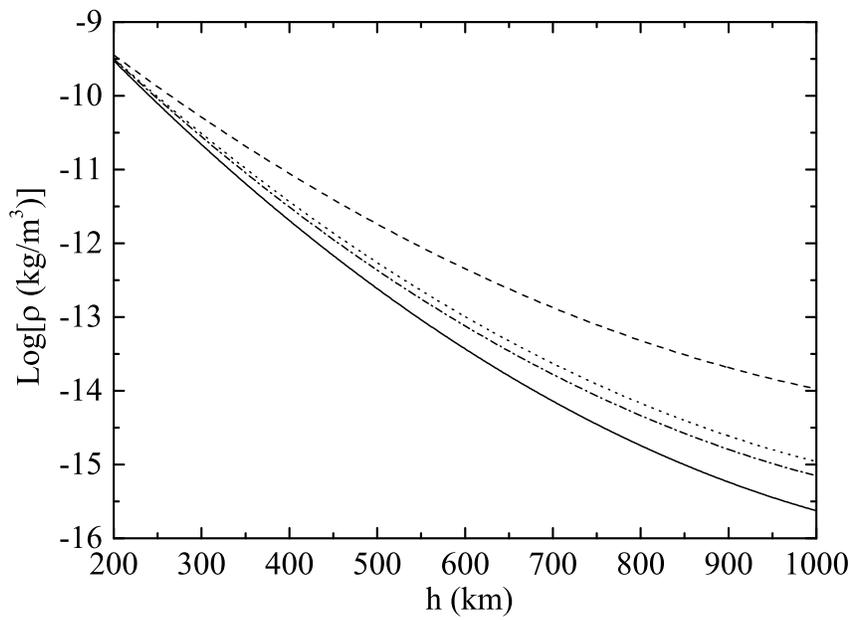}
\caption{Decimal logarithm of the atmospheric density (in Kg$/$m$^3$ vs altitude in kilometers during the Galileo I (dashed line), Galileo II (dotted line), Juno (dashed-dotted line) and NEAR flybys (solid line).}
\label{fig:1}       
\end{figure}

In Fig.\ \ref{fig:1} we have plotted the density profiles obtained from Eqs. (\ref{temp})-(\ref{profile}) and the data in
table \ref{tab1}. We see that the density was higher on December, 8th, 1990 during the first flyby of the Galileo spacecraft
because it coincided with a maximum of Solar activity roughly at the peak of the sunspot cycle $22$. On the contrary, the NEAR
flyby occurred at the minimum between solar cycles number $22$ and $23$, so we see that the density of the thermosphere was at
a minimum on that event.

The drag force \cite{KingHele,DragSphere} acts opposite to the spacecraft velocity vector and it is proportional to the square of the magnitude of the
velocity, $v$, as follows:

\begin{equation}
\label{drag}
\mathbf{D}=-\displaystyle\frac{1}{2} \, \rho \, v^2 \, A\, C_d \, \mathbf{\hat{v}} \; ,
\end{equation}

where $\rho$ is the atmospheric density, $A$ is the cross-sectional area perpendicular to the velocity, $C_d$ is a drag
coefficient (usually taken as $C_d=2$) and $\mathbf{\hat{v}}$ is the unit vector in the direction of the velocity.
We also notice that some recent works approach this as an inverse problem by deducing the atmospheric density from the
friction exerted upon satellites without reference to any particular atmospheric model \cite{Picone}. However, we have no direct information on the density profile from direct measurements at the time of the aforemention flybys so we will use the model discussed in this section.

Besides the drag component of the perturbing force we should also consider a lift component:

\begin{equation}
\label{lift}
\mathbf{L}=\displaystyle\frac{1}{2} \, \rho\, v^2 \, S \, C_l \, \mathbf{\hat{h}}\; ,
\end{equation}

where $S$ is the area of the wings (in the case of aircraft) or any surface acting as such and $C_l$ is the lift coefficient. Here, $\mathbf{\hat{h}}$ is a unit vector perpendicular to the spacecraft velocity \cite{Clancy}. As the coefficient $C_l$ is 
a quantity of order one we have that, in magnitude, the effect of the lift and the drag forces are similar. However, as 
$\mathbf{L}$  acts perpendicularly to the spacecraft velocity we have its effect on the variation of the magnitude of the total
velocity of the spacecraft is much smaller than that of $\mathbf{D}$. So, in this paper we will only consider the drag component of the force in Eq. (\ref{drag}). Furthermore, spacecraft are not designed as aerodynamic vehicles and we expect smaller values
of $C_l$ than those corresponding to aircraft.

We must also notice that our atmospheric model uses daily averages for the Solar activity parameters $F10.7$ and $A_p$. Although the solar radio flux varies according to the Solar cycles and it is relatively stable throughout the day, the geomagnetic index, $A_p$, may show variations around $\pm 5$, specially in the case of Solar storms \cite{Du,Muller}.  This corresponds to an uncertainty
in the thermospheric density given by:

\begin{equation}
\label{deltarho}
\displaystyle\frac{\delta \rho}{\rho} \simeq (h-175) m \displaystyle\frac{\delta T}{T} \; ,
\end{equation}

where $\delta T$ gives an estimate on the changes of space weather. For example, for the Galileo II flyby we have that $T=1086.5$
${}^\circ K$ and from the data in Table \ref{tab1} and Eqs. (\ref{temp}), (\ref{mol}) and (\ref{deltarho}) we have that
for $\delta A_p=5$ the uncertainty in the thermosphere's temperature is $\delta T=7.5$ ${}^\circ K$ and, consequently, the
uncertainty in the density is around a $6\, \%$. The variations due to the day-night cycle are more important as they can change the
thermospheric density by a factor $2$ in some days, with an average variation of $1.5$, approximately \cite{Muller}. We are not
taking into account these fluctuations in our model but the effects on the predicted change of the spacecraft velocity magnitude
would be, at most, a fifty per cent. On the other hand, we should mention that the Galileo II flyby (for which the largest friction effect was detected) took place on the daylight hemisphere because its perigee was attained at the latitude
$354.4^\circ$ at $15:09$ UTC \cite{AcedoMNRAS}.

In the next section we will calculate the perturbation induced by the dragging force in Eq. (\ref{drag}) on the trajectories of the spacecraft listed in table \ref{tab1} by using the atmospheric model in Eqs. (\ref{temp})-(\ref{profile}).

\section{Perturbation of the osculating orbit at perigee}
\label{pertur}

The contribution of small forces to the deformation of trajectories in celestial mechanics is traditionally studied
using perturbation theory \cite{Danby,Burns}. In this approach we consider the osculating orbit at a given point and we solve the system
of equations of motion for the perturbations around the point of interest. This osculating orbit is defined as the ideal
keplerian orbit that the spacecraft should follow in case all the perturbations are switched off at that point so this orbit
is tangential to the real orbit in phase space just at the point considered. 

This technique is also particularly well-suited to our problem because we are interested in the effect of atmospheric friction
in the short period of time that the spacecraft spends at the thermosphere and the low exosphere, i. e., for altitudes between
the perigee to $1000$-$1200$ km over the surface of the Earth. For this reason, we will define the osculating orbit at
perigee (whose semi-major axis, $a$, and eccentricity, $\epsilon$, can be calculated from the ephemeris) at the perigee and another point close to the perigee (one or five minutes after the perigee). These are obtained by solving the following system of 
equations \cite{Danby,Burns}:

\begin{eqnarray}
\label{osculating}
r_t &=& r_P \displaystyle\frac{\epsilon \cosh \eta-1}{\epsilon-1} \\
\noalign{\smallskip}
t &=& \sqrt{\displaystyle\frac{r_P^3}{\mu}}\displaystyle\frac{\left(\epsilon \sinh \eta-\eta\right)}{\left( \epsilon-1\right)^{3/2}} \; ,
\end{eqnarray}
where $r_P=R_E+h$ is the minimum distance to the center of the Earth attained by the spacecraft during the flyby given
by the sum of the Earth radius at the location of the vertical of the perigee, $R_E$, and the altitude, $h$, of the spacecraft. Here $r_t$ is the distance to the center of the Earth at time $t$ after the perigee, $\epsilon$ is the orbital eccentricity, $\eta$ is the eccentric anomaly for the osculating orbit at time $t$ and $\mu= G M_E =398600.4$ km$^3/$s$^2$ is the product of the
gravitational constant and the mass of the Earth.

Once we have obtained the eccentricity, the semi-major axis, $a$, is deduced from the identity $r_P=a (1-\epsilon)$. By 
considering $t=300$ seconds in Eq. (\ref{osculating}) we have obtained the eccentricity and the semi-major axis listed
in table \ref{tab2} for each flyby. The cross product of the position vectors at perigee and time $t$ also give us the
inclination vector from which we can determine the orientation of the orbital plane. The rest of necessary parameters 
are deduced either directly from the ephemeris or from standard relations in celestial mechanics. 

\begin{table*}
\caption{Parameters for the osculating orbit at perigee for some spacecraft flybys of the Earth: $\theta_{\textsc{P}}$, $\alpha_{\textsc{P}}$ are the polar angle and the right ascension for the direction of the perigee, $I$ and $\alpha_{\textsc{I}}$ are the polar angle and the right ascension for the inclination vector, $V_{\textsc{P}}$ and $V_{\infty}$ are the magnitude of the velocities at perigee and at a large distance from the center of the Earth (asymptotic velocity). \label{tab2}}
\centering
\resizebox{\textwidth}{!}{
 \begin{tabular}{lcccccccc}
  Spacecraft & $a$ (km) & $\epsilon$ (km) & $\theta_{\textsc{P}}$ & $\alpha_{\textsc{P}}$ & $I$ 
 & $\alpha_{\textsc{I}}$ & $V_{\textsc{P}}$ (km$/$s) & $V_{\infty}$ (km$/$s) \\
    \noalign{\smallskip}
   Galileo I & -4978.17 & 2.4726 & 66.04$^\circ$ & 319.96$^\circ$ & 142.9$^\circ$ & 13.97$^\circ$ & 13.74 & 8.949  \\
  \noalign{\smallskip}
   Galileo II & -5059.26 & 2.3191 & 123.8$^\circ$ & 302.72$^\circ$ & 138.7$^\circ$ & 82.36$^\circ$ & 14.08 & 8.877  \\
  \noalign{\smallskip} 
   NEAR & -8494.87 & 1.81352 & 57.0$^\circ$ & 280.42$^\circ$ & 108.0$^\circ$ & 358.24$^\circ$ & 12.74 & 6.850  \\
  \noalign{\smallskip}
   Juno & -4139.7 & 2.6740 & 123.39$^\circ$ & 344.13$^\circ$ & 47.13$^\circ$ & 291.85$^\circ$ & 14.53 & 9.813  
  \end{tabular}}
\end{table*}

It is also convenient to perform the calculations in an orbital frame of reference defined by three unit vectors:
one of them along the position vector at perigee, $\mathbf{\hat{s}}$, a second one along the orbital inclination
vector, $\mathbf{\hat{w}}$ and a third one, $\mathbf{\hat{n}}$, obtained as the cross product of $\mathbf{\hat{s}}$ and $\mathbf{\hat{w}}$ in such a way that its direction coincides with that of the velocity vector at perigee \cite{Acedo2014}.  Both 
$\mathbf{\hat{s}}$ and $\mathbf{\hat{w}}$ can be obtained from the polar and right ascension angles in table \ref{tab2}.

The osculating hyperbolic keplerian orbit at perigee is then defined by the equations:

\begin{eqnarray}
\label{Dkepler}
\mathbf{R}&=&-\left(\cosh \eta-\epsilon\right)\, \mathbf{\hat{s}}+\sqrt{\epsilon^2-1} \sinh \eta\, \mathbf{\hat{n}} \; ,\\
\noalign{\smallskip}
\label{Vkepler}
\mathbf{V}&=&\displaystyle\frac{\sinh \eta \, \mathbf{\hat{s}}-\sqrt{\epsilon^2-1} \cosh \eta \, \mathbf{\hat{n}} }{\epsilon\cosh \eta-1} \; ,
\end{eqnarray}
where $\eta$ is the eccentric anomaly \cite{Burns,Acedo2014}. Notice that the distance is measured in units of $\vert a \vert$ and the velocity
in units of $\vert a \vert/T$, $T=\sqrt{\vert a \vert^3/\mu}$ being the characteristic time-scale for the keplerian orbit.

The perturbed orbit is then given by $\mathbf{R}+\delta \mathbf{r}$, $\mathbf{V}+\delta \mathbf{v}$, where $\vert \delta r \vert$ and $\vert \delta v \vert$ are small compared to $\vert \mathbf{R} \vert$ and $\vert \mathbf{V} \vert$, respectively. By considering small variations in the Newtonian equation of motion we get:
\begin{eqnarray}
\label{eqmotionr}
\displaystyle\frac{d \delta \mathbf{r}}{d \tau} &=& \delta  \mathbf{v} \; , \\
\noalign{\smallskip}
\label{eqmotionv}
\displaystyle\frac{d \delta \mathbf{v}}{d \tau} &=& \displaystyle\frac{\mathbf{D}}{m} \displaystyle\frac{T^2}{\vert a \vert}-
\displaystyle\frac{\delta \mathbf{r}}{R^3}+3 \mathbf{R}\cdot \delta \mathbf{r} \displaystyle\frac{\mathbf{R}}{R^5} \; ,
\end{eqnarray}
where we have included the factor $T^2/\vert a \vert$ in the friction acceleration to cast it also into non-dimensional form and $\tau=t/T$. These equations can be numerically integrated by any standard numerical method, such as fourth-order Runge-Kutta, with
the initial conditions $\delta \mathbf{r}(\tau=0)=\mathbf{0}$, $\delta \mathbf{v}(\tau=0)=\mathbf{0}$. The non-dimensional time
can also be written in terms of the eccentric anomaly using:
\begin{equation}
\label{time}
\tau=\epsilon \sinh\eta-\eta\; ,
\end{equation}
and, as $\mathbf{R}$ and $\mathbf{V}$ in Eqs. (\ref{Dkepler}) and (\ref{Vkepler}) are more clearly related with the eccentric
anomaly, $\eta$, we will integrate the system in Eqs. (\ref{eqmotionr}) and (\ref{eqmotionv}) using $\eta$ instead of 
non-dimensional time, $\tau$. In the next section we will integrate the system of equations of motion in Eqs. (\ref{eqmotionr}) and (\ref{eqmotionv}) to obtain a bound on the kinematic effect of friction in four flybys whose perigee was attained at the
Earth's exosphere.

\section{Numerical results}
\label{results}

To integrate the equations of motion we finally need an estimation of the cross-sectional area of the spacecraft during
the flyby as it appears in Eq. (\ref{drag}). The geometry of spacecraft varies from mission to mission according to the
scientific instruments to be transported and the size and distribution of the solar panels or radioisotope
thermoelectric generators (RTGs) required for producing the electricity necessary to power those instruments. The absence of any
relevant friction in space allows for a large design freedom not possible for aircraft. This also increases the difficulty
to obtain a good estimation of the spacecraft area as it crosses the thermosphere during a flyby. For that reason, we will
provide some reasonable upper and lower estimates for the Galileo mission (because this experiences the larger friction in
his second flyby) and upper bounds for the NEAR and Juno spacecraft.

Anyway, we must also emphasize that the orientation of the different parts of the spacecraft and the resulting shadowing and
reflection effects might also have an influence which is not covered by our effective surface model. Moreover, in the drag
model of Eq. (\ref{drag}), we are considering that $C_d=2$ which, in fact, represents a worst case scenario. This coefficient
depends upon the materials and the shape of the spacecraft and it could differ largely from the assumed value. Experiments
performed for smooth and rough spheres in a wind tunnel for a wide range of Reynold numbers have shown that $0.1 \le C_d \le 1.5$ \cite{DragSphere}. As no results from wind tunnel experiments for these spacecraft are reported we will assume $C_d \simeq 2$ to obtain upper bounds for the dragging effect on the velocity variation during the flyby. Notwithstanding the lack of experimental
data for the dragging coefficients of interplanetary spacecraft we should mention that some studies of aeronomic satellites composed by cylindrical surfaces and flat plates suggest that $C_d \gtrsim 2$ and this could be adequate for estimating this
parameter for the NEAR, Galileo or Juno spacecraft \cite{Moe}.

\begin{figure}
\includegraphics[width=\columnwidth]{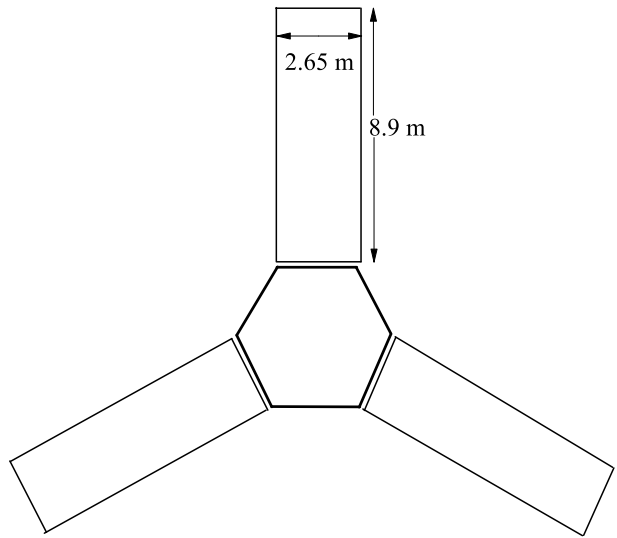}
\caption{A top view of the Juno spacecraft. The main body is a prism with hexagonal base but the dominant feature are
the three solar panels with a length of $8.9$ meters and a withd of $2.65$ meters. These large panels provide the necessary energy as the spacecraft is orbiting Jupiter.}
\label{fig:2}       
\end{figure}

\begin{figure}
\includegraphics[width=\columnwidth]{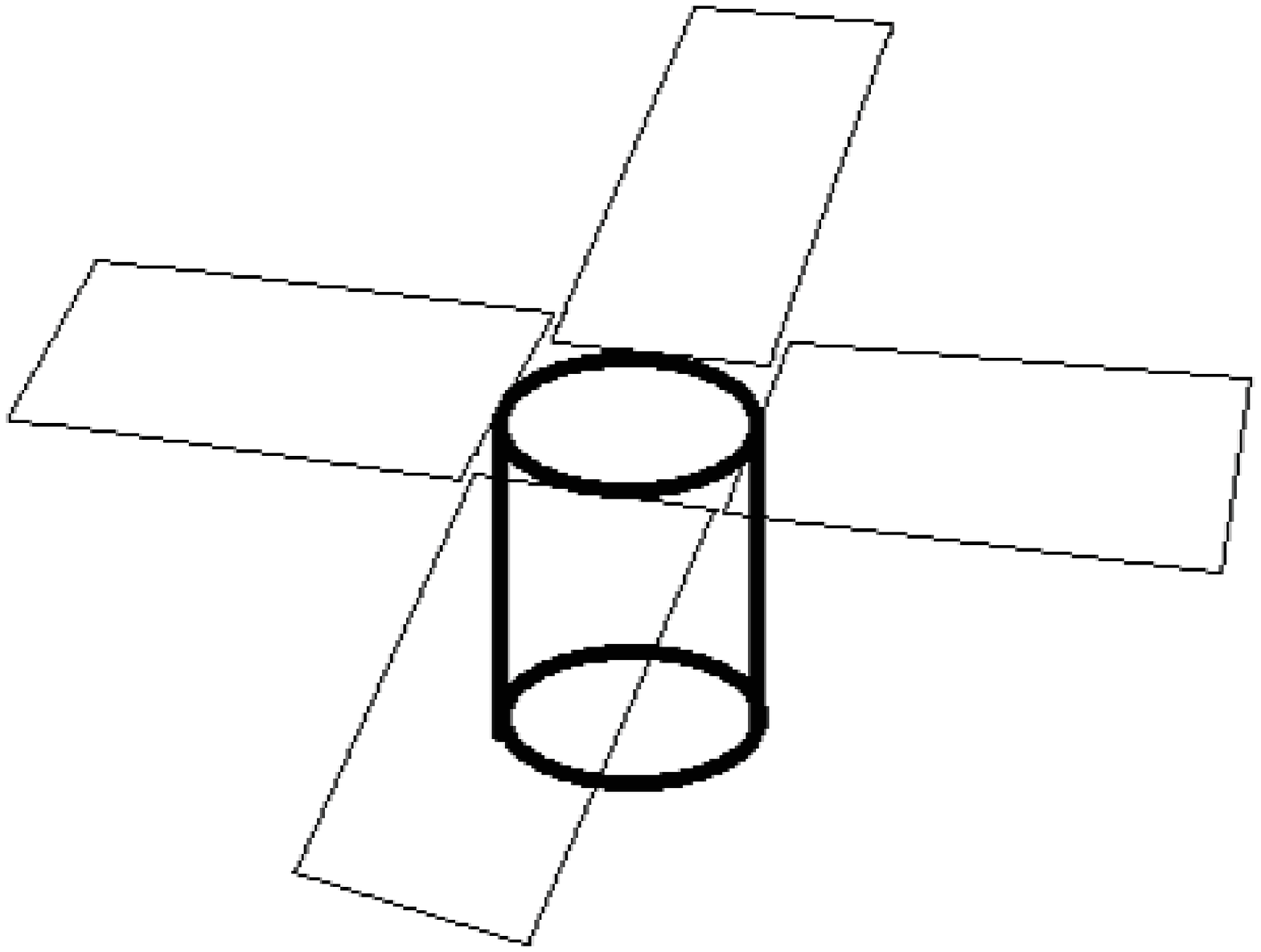}
\caption{The NEAR spacecraft is shaped as a cylinder $1.7$ meters high with four solar panels whose area is $1.5 \times  2.75$ square meters.}
\label{fig:3}       
\end{figure}

The Galileo spacecraft measured $5.3$ meters from the low gain antenna to the atmospheric probe designed to descent into
Jupiter's atmosphere \cite{Galileo}. Its maing geometric features were the sun shields enveloping the low gain antenna, the RTGs projecting
from the main body and a large shaft mounting two magnetometer sensors and a plasma-wave antenna at its tip. But, on the whole,
it was a compact object whose cross-section, depending on the orientation, can be estimated in the range $4^2 \mbox{ m$^2$} < A < 5.3^2 \mbox{ m$^2$}$. On the other hand, Juno is shaped as a hexagonal prism with three large solar panels regularly attached
\cite{JunoMissionI,JunoMissionII,JunoMissionIII}. This gives it a span of $20$ meters as sketched in Fig. \ref{fig:2}. The NEAR spacecraft was composed by a cylindrical structure
and four solar panels smaller than those of Juno because it was designed to operate in the inner solar system where radiation
is more intense \cite{NEAR}. The sketch is shown in Fig. \ref{fig:3}.

\begin{figure}
\includegraphics[width=\columnwidth]{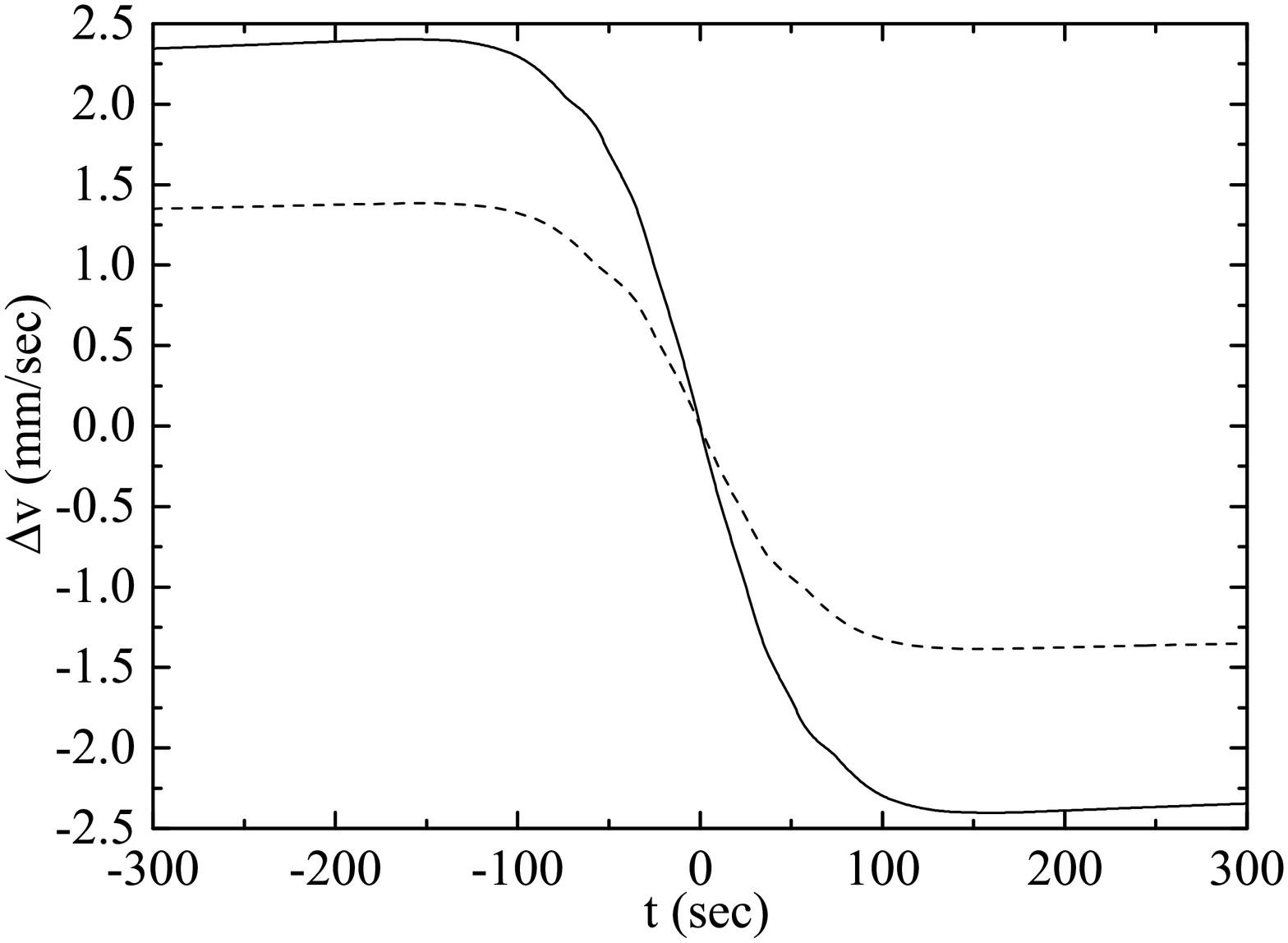}
\caption{Velocity modulus perturbation for the Galileo II flyby of Earth as a consequence of friction in mm per second vs time in seconds. The solid line correspond to a cross-sectional area estimate of $5.3^2$ square meters. The dotted line corresponds to the lower estimate of $4^2$ square meters. The drag coefficient was taken as $C_d=2$, as usual in satellite calculations.}
\label{fig:4}       
\end{figure}

By integrating the system of perturbation equations of motions in Eqs. (\ref{eqmotionr}) and (\ref{eqmotionv}), with the drag
force in Eq. (\ref{drag}), the mass in Table \ref{tab1} and the dimensions given above we obtain an estimation of the
perturbation in the velocity magnitude defined as:
\begin{equation}
\label{vpert}
\Delta v = \vert \mathbf{V}+\delta \mathbf{v}\vert-\vert \mathbf{V} \vert \; ,
\end{equation}
where $\mathbf{V}$ is the velocity corresponding to the unperturbed osculating orbit at perigee as given in Eq. \ (\ref{Vkepler}). The result for the Galileo II flyby, as a function of time since the perigee, is shown in Figure \ref{fig:4}. The upper and the lower curve are the predictions for the estimated areas of $5.3^2$ m$^2$ and $4^2$ m$^2$, respectively.

Integration was performed from $\eta=-0.4$ until $\eta=0.4$, which corresponds, according to Eq. (\ref{time}), to a period
of $314.94$ seconds, before and after the perigee. From the results in Fig. \ref{fig:4}, we conclude that atmospheric friction
contributed to the deceleration of the Galileo spacecraft with a value in the range:
\begin{equation}
\label{GalII}
-4.67 \mbox{ mm$/$s} < \Delta v < -2.69 \mbox{ mm$/$s}\; ,
\end{equation}
and, as a consequence, we cannot explain completely the detected anomalous velocity decrease of $-8$ mm$/$s. Anderson et al.
\cite{Anderson2008} quoted in their paper that an estimate of $\Delta v=-3.4$ mm$/$s (for the atmospheric friction effect) which coincides with our calculations. The influence of
atmospheric drag in the other flybys was far smaller because their perigees were as a higher altitude from the surface of the
Earth. For the first flyby of the Galileo in 1990 we obtain that it was below the altitude of $1000$ km for a period of 
$331.35$ seconds, a similar calculation using the atmospheric model depicted in Fig. \ref{fig:1} yields a perturbation
range:
\begin{equation}
\label{GalI}
-4.28\times 10^{-3} \mbox{ mm$/$s} < \Delta v < -2.44 \times 10^{-3} \mbox{ mm$/$s} \; .
\end{equation}
This is totally undetectable in comparison with the anomalous velocity increase of $3.92$ mm$/$s reported by Anderson et al.
\cite{Anderson2008}.

We have also that the NEAR spacecraft spent, approximately, $614$ seconds below the critical altitude of the atmospheric model, 
$h < 1200$ km. For the Juno spacecraft this period lasted for $495$ seconds. If we now use the masses and atmospheric indexes as listed in Table \ref{tab1}, the maximum areas corresponding to the solar panels as shown in Figs. \ref{fig:2} and \ref{fig:3} and the atmospheric model of Eq. (\ref{profile}) we obtain the following upper bounds for the velocity perturbation:

\begin{eqnarray}
\label{NEARdelta}
\vert \Delta v \vert &<& 0.013 \mbox{ mm$/$s for the NEAR flyby}\; , \\
\noalign{\smallskip}
\label{Junodelta}
\vert \Delta v \vert &<& 0.04 \mbox{ mm$/$s for the Juno flyby}\; ,
\end{eqnarray}

which are small but on the threshold of detectability in the Doppler residuals. We should also remember that an anomalous
velocity increase of $13.46$ mm$/$s was found for the NEAR flyby \cite{Anderson2008} and that no anomaly was reported after the orbit reconstruction
for the Juno flyby \cite{Thompson}. Notice that Thompson et al. \cite{Thompson} have also reported a maximum effect of $0.1$ mm$/$s after turning off the atmospheric drag in their orbital model, in agreement with our independent estimation. 

\begin{table*}
\caption{Results for the modelling of the variation in the magnitude of the spacecraft velocity as a consequence
of atmospheric friction compared with the anomalies detected. \label{tab3}}
\centering
\resizebox{\textwidth}{!}{
 \begin{tabular}{lccccc}
  Spacecraft flyby & Area in m$^2$ & Observed $\Delta V_\infty$ (mm$/$sec) & Predicted $\Delta V_\infty$ (mm$/$sec)  \\
  \noalign{\smallskip}
  Galileo II & $22.04 \pm 6.04$ & -8 & $-3.68\pm 0.99$ \\
  \noalign{\smallskip}
  Galileo I & $22.04\pm 6.04$ & 3.92 &  $(-3.36\pm 0.92)\times 10^{-3}$\\
  \noalign{\smallskip}
  NEAR & $16.5$ (maximum) & 0 & $0.04$ \\
  \noalign{\smallskip}
  Juno & $70.75$ (maximum) & 13.46 & $0.013$
  \end{tabular}}
\end{table*}

The results for the four flybys we have analyzed are summarized on Table \ref{tab3}.

\begin{figure}
\includegraphics[width=\columnwidth]{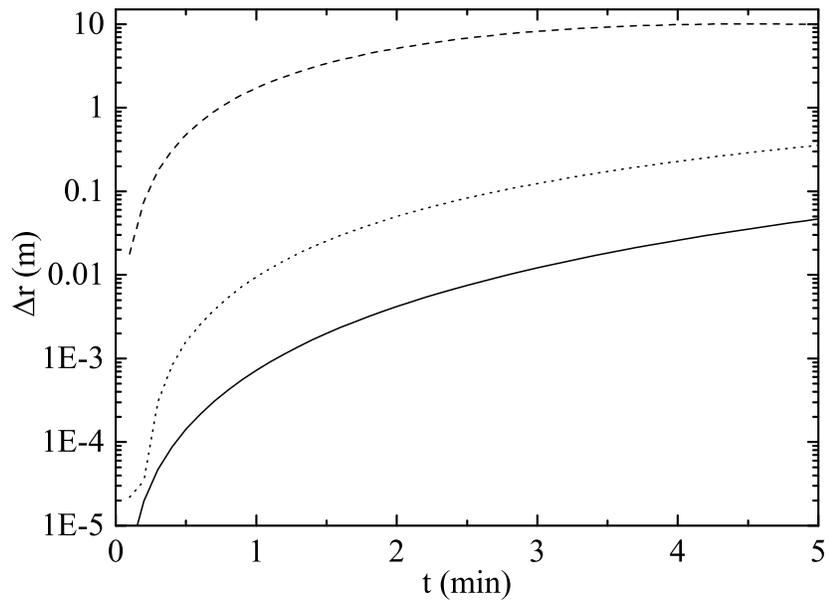}
\caption{Perturbations in the radial coordinate (in metres) for the Galileo spacecraft in its second flyby of the Earth. The dashed line corresponds to the contribution of the zonal and tesseral harmonics in the EGM96 geopotential model, the dashed line to the friction perturbation and the solid line to the tidal forces induced by the Sun and the Moon. Time is measured in minutes since the perigee.}
\label{fig:5}       
\end{figure}

Finally, it would be also interesting to analyze the perturbations in the radial coordinate (the distance to the center of the Earth) as a consequence of atmospheric friction as deduced from the integration of the system of equations of motion in Eqs. (\ref{eqmotionr}) and (\ref{eqmotionv}). In Fig. \ref{fig:5} we plot the results for the period of time in which the Galileo spacecraft remained in the thermosphere in comparison with other sources of perturbation: the tidal forces by the Sun and the Moon \cite{Acedo2015} and the contributions from the corrections to the Earth's potential in the EGM96 model \cite{EGM96,AcedoMNRAS}.

Notice that the most important effect around the perigee is that of the zonal and tesseral harmonics which imply a deviation
around $10$ metres from the osculating Keplerian orbit at perigee. The atmospheric friction effect amounts to a perturbation of
several decimetres in the spacecraft position and it would be just in the threshold of detectability in the modern 
applications of the Delta Differential One-Way Ranging for the Deep Space Network \cite{DeltaDOR}. It is possible that in future missions a detection of this effect on the perturbed orbit during the flyby could be achieved \cite{NanoRad}.

\section{Conclusions and Remarks}
\label{conclusions}

Atmospheric drag is an important effect to be considered for satellites orbiting inside the exosphere of planets. Even for the
LAGEOS satellite, currently orbiting as an almost circular orbit at an altitude of $5900$ km, there is a small contribution
to its decay associated to atmospheric drag from neutral hydrogen, the major part corresponding to charge drag \cite{Rubincam}. This effect
is more important for satellites that for interplanetary spacecraft performing flybys because its effect is cumulative in the first case but the flyby
is an isolated event in the interplanetary spacecraft mission. Nevertheless, the contribution of atmospheric friction to the
perturbation of the trajectories of spacecraft during a flyby can now be detected with the improvement of Doppler tracking
techniques and orbital models \cite{Anderson2008,Thompson}.

Moreover, the recent discovery of some anomalous velocity changes in spacecraft flybys has stimulated the interest in this
problem, apart from its intrinsic application to spacecraft dynamics in interplanetary missions \cite{KingHele}. In this paper we have
applied simple drag and atmospheric models to estimate the velocity perturbation in several missions in which the
spacecraft perigee was attained at a point inside the thermosphere. The larger velocity decrease in any of these missions in the
last decades was achieved by the second Galileo flyby around the Earth on December, 8th, 1992. After fitting the orbit, engineers
at NASA found a total anomalous velocity decrease of $-8$ mm$/$s as revealed in the post-encounter Doppler residuals. Anderson et al. \cite{Anderson2008}
pointed out that $-3.4$ mm$/$s was expected from atmospheric friction because the minimum altitute of the spacecraft in this
event was $303$ km (a point lying, approximately, in the middle layer of the thermosphere).

We have evaluated this kinematic effect by considering possible values of the spacecraft cross-sectional area during the 
flyby and our results back up those reported by Anderson et al. \cite{Anderson2008}. Consequently, we must conclude that there still remains 
an explained residual decrease of several mm$/$s in the Galilleo II flyby after substracting the effect of atmospheric friction. 
A complete explanation of the anomalous decrease in terms of atmospheric friction would require a larger drag coefficient, $C_d$, 
than usually considered, or an enhanced friction as a consequence of eddy viscosity in turbulent flows \cite{MoninYaglom}. 
Nevertheless, experiments performed at NASA with smooth and round spheres have shown that the drag coefficient is smaller than
$C_d < 1.5$ for a broad range of Reynold numbers \cite{DragSphere}. No experiments for the more complicate geometry of the Galileo spacecraft have
been reported but taken $C_d \simeq 2$ can be considered as a reasonable upper bound in view of other experiments on wind tunnels and theoretical models \cite{Moe}.

For flybys with perigees at altitudes of $500$ km or higher we have shown that only a decrease of a fraction of mm$/$s in the
post-encounter velocity can be expected as a consequence of friction alone. This is also consistent with other orbit reconstruction studies \cite{Thompson}. So, we can fairly say that no significant contribution to the final velocity of the NEAR flyby
can be deduced from atmospheric friction and that it can be ruled out as a possible explanation of the anomalous velocity
change detected in that case.

From the point of view of planetary science, orbital models for the Juno spacecraft flyby of Jupiter could provide
information about its exosphere if similar orbital determination techniques are applied there after the retrieval of the
mission data.

%
%

\bibliographystyle{plain}      

\bibliography{acedobiblio}

  
%
%

\end{document}